\documentclass[english,superscriptaddress]{elsarticle}
\usepackage[T1]{fontenc}
\usepackage[latin9]{inputenc}
\usepackage{geometry}
\usepackage{amsthm}
\usepackage{amsmath}
\usepackage{amssymb}
\usepackage{graphicx}
\usepackage{esint}

\makeatletter
\numberwithin{equation}{section}
\numberwithin{figure}{section}

\usepackage{slashed}
\journal{CR Physique}

\makeatother

\usepackage{babel}
\begin{document}

\title{Recent developments in transport phenomena in Weyl semimetals}

\author{Pavan Hosur}

\address{Department of Physics, Stanford University, Stanford, CA 94305, USA}

\author{Xiaoliang Qi}

\address{Department of Physics, Stanford University, Stanford, CA 94305, USA}
\begin{abstract}
The last decade has witnessed great advancements in the science and
engineering of systems with unconventional band structures, seeded
by studies of graphene and topological insulators. While the band
structure of graphene simulates massless relativistic electrons in
two dimensions, topological insulators have bands that wind non-trivially
over momentum space in a certain abstract sense. Over the last couple
of years, enthusiasm has been burgeoning in another unconventional
and topological (although, not quite in the same sense as topological
insulators) phase -- the Weyl Semimetal. In this phase, electrons
mimic Weyl fermions that are well-known in high-energy physics, and
inherit many of their properties, including an apparent violation
of charge conservation known as the Chiral Anomaly. In this review,
we recap some of the unusual transport properties of Weyl semimetals
discussed in the literature so far, focusing on signatures whose roots
lie in the anomaly. We also mention several proposed realizations
of this phase in condensed matter systems, since they were what arguably
precipitated activity on Weyl semimetals in the first place.
\end{abstract}
\maketitle

\section{Introduction to Weyl semimetals}

The earliest classification of the forms of matter in nature, typically
presented to us in our early school days, consists of \emph{solids},
\emph{liquids} and \emph{gases}. High school physics textbooks and
experience later teach us that solids can be further classified based
on their electronic properties as \emph{conductors} and \emph{insulators}.
They tell us that as long as the electrons in a solid are non-interacting,
solids with partially filled bands are metals or conductors while
those with no partially filled bands and a gap between the valence
and the conduction bands are insulators or semiconductors. Solid state
physics courses in college add another phase to this list: if the
gap is extremely small or vanishing, or if there is a tiny overlap
between the valence and the conduction bands, the material is \emph{semimetallic}
and has markedly different electronic properties from metals and insulators.
Graphene (\citet{GeimGraphene}) -- a two dimensional (2D) sheet of
carbon atoms -- is the most celebrated example of a semimetal with
a vanishing gap. In this system, the conduction and valence bands
intersect at certain points in momentum space known as Dirac points.
The dispersion near these points is linear and electrons at nearby
momenta act like massless relativistic particles, thus stimulating
the interest of condensed matter and high-energy physicists alike.

In the last couple of years, there has been growing interest in a
seemingly close cousin of graphene -- the so-called \emph{Weyl semimetal}
(WSM) (\citet{PyrochloreWeyl,KrempaWeyl,ChenIridate,TurnerTopPhases,VafekDiracReview,VolovikBook,VolovikFlatBands}).
Like graphene, its band structure has a pair of bands crossing at
certain points in momentum space; unlike graphene, this is a three-dimensional
(3D) system. Near each such \emph{Weyl point} or \emph{Weyl node},
the Hamiltonian resembles the Hamiltonian for the Weyl fermions that
are well-known in particle physics literature:
\begin{equation}
H_{\mbox{Weyl}}=\sum_{i,j\in\{x,y,z\}}\hbar v_{ij}k_{i}\sigma_{j}\label{eq:Weyl Hamiltonian}
\end{equation}
where $\hbar$ is the reduced Planck's constant, $v_{ij}$ have dimensions
of velocity, $k_{i}$ is the momentum relative to the Weyl point and
$\sigma_{i}$ are Pauli matrices in the basis of the bands involved.
Thus, the name \emph{Weyl semimetal}.

A closer look, however, unveils a plethora of differences between
graphene and WSMs because of their different dimensionality. An immediate
consequence of the form of $H_{\mbox{Weyl}}$ is that the Weyl points
are topological objects in momentum space. Since all three Pauli matrices
have been used up in $H_{\mbox{Weyl}}$, there is no matrix that anticommutes
with $H_{\mbox{Weyl}}$ and gaps out the spectrum. There are then
only two ways a Weyl point can be destroyed. The first is by annihilating
it with another Weyl point of opposite chirality, either by explicitly
moving the Weyl points in momentum space and merging them or by allowing
for scattering to occur between different Weyl nodes; the latter requires
the violation of translational invariance. The second way is by violating
charge conservation via superconductivity. Thus, given a band structure,
the Weyl nodes are stable to arbitrary perturbations as long as charge
conservation and translational invariance is preserved. Disorder,
in general, does not preserve the latter symmetry; however, if the
disorder is smooth, many properties of the WSM that rely on the topological
nature of the band structure should survive. In contrast, Dirac nodes
in graphene can be destroyed individually by breaking lattice point
group symmetries.

The topological stability of the Weyl nodes crucially relies on the
intersecting bands being non-degenerate. For degenerate bands, terms
that hybridize states within a degenerate subspace can in general
gap out the spectrum. Thus, the WSM phase necessarily breaks at least
one out of time-reversal and inversion symmetries, as the presence
of both will make each state doubly degenerate.

Based on (\ref{eq:Weyl Hamiltonian}), each Weyl point in a WSM can
be characterized by a \emph{chirality quantum number} $\chi$ defined
as $\chi=\mbox{sgn}[\mbox{det}(v_{ij})]$. The physical significance
of the chirality is as follows. An electron living in a Bloch band
feels an effective vector potential $\boldsymbol{A}(\boldsymbol{k})=i\left\langle u(\boldsymbol{k})|\boldsymbol{\nabla_{k}}u(\boldsymbol{k})\right\rangle $
because of spatial variations of the Bloch state $|u(\boldsymbol{k})\rangle$
within a unit cell. The corresponding field strength $\boldsymbol{F}(\boldsymbol{k})=\boldsymbol{\nabla}_{\boldsymbol{k}}\times\boldsymbol{A}(\boldsymbol{k})$,
known as the Berry curvature or the Chern flux, plays the role of
a magnetic field corresponding to the vector potential $\boldsymbol{A}(\boldsymbol{k})$.
It can be shown that $\boldsymbol{F}(\boldsymbol{k})$ near a Weyl
node of chirality $\chi$ satisfies 
\begin{equation}
\frac{1}{2\pi}\oint_{FS}\boldsymbol{F}(\boldsymbol{k})\cdot\mathrm{d}\boldsymbol{S}(\boldsymbol{k})=\chi\label{eq:monopole chirality}
\end{equation}
where the integral is over any Fermi surface enclosing the Weyl node
and the area element $\mathrm{d}\boldsymbol{S}(\boldsymbol{k})$ is
defined so as to point away from the occupied states. Since $\boldsymbol{F}(\boldsymbol{k})$
acts like a magnetic field in momentum space, (\ref{eq:monopole chirality})
suggests that a Weyl node acts like a magnetic monopole in momentum
space whose magnetic charge equals its chirality. Equivalently, a
Fermi surface surrounding the Weyl node is topologically non-trivial;
it has a Chern number, defined as the Berry curvature integrated over
the surface as in (\ref{eq:monopole chirality}), of $\chi$ ($-\chi$)
for an electron (a hole) Fermi surface. 

\begin{figure}
\begin{centering}
\includegraphics[width=0.5\columnwidth]{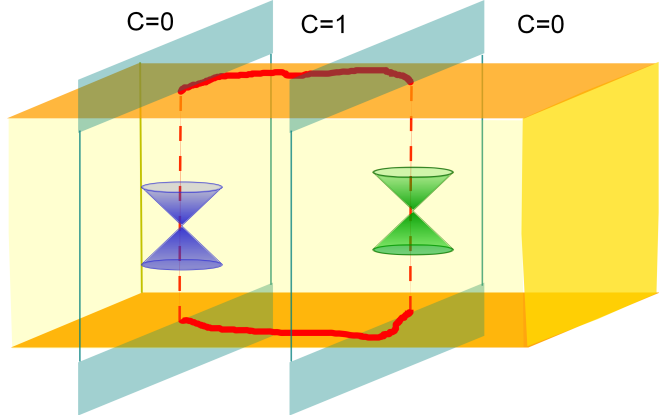}
\par\end{centering}

\caption{Weyl semimetal with a pair of Weyl nodes of opposite chirality (denoted
by different colors green and blue) in a slab geometry. The surface
has unusual Fermi arc states (shown by red curves) that connect the
projections of the Weyl points on the surface. $C$ is the Chern number
of the 2D insulator at fixed momentum along the line joining the Weyl
nodes. The Fermi arcs are nothing but the gapless edge states of the
Chern insulators strung together. \label{fig:Weyl-semimetal}}
\end{figure}

\citet{NielsenFermionDoubling1,NielsenFermionDoubling2} showed that
the total magnetic charge in a band structure must be zero, which
implies that the total number of Weyl nodes must be even, with half
of each chirality. The argument is simple and runs as follows. Each
2D slice in momentum space that does not contain any Weyl nodes can
be thought of as a Chern insulator. Since Weyl nodes emit Chern flux,
the Chern number changes by $\chi$ as one sweeps the slices past
a Weyl node of chirality $\chi$. Clearly, the Chern numbers of slices
will be periodic across the Brillouin zone if and only if there are
as many Weyl nodes of chirality $\chi$ as there are of chirality
$-\chi$. Such a notion of chirality does not exist for graphene or
the surface states of topological insulators, which also consist of
2D Dirac nodes, because the Berry phase around a Fermi surface is
$\pi$ which is indistinguishable from $-\pi$.

The fact that each Weyl node is chiral and radiates Chern flux leads
to another marvelous phenomenon absent in two dimensions -- the \emph{chiral
anomaly}. The statement is as follows: suppose the universe (or, for
condensed matter purposes, the band structure) consisted only of Weyl
electrons of chirality $\chi$ and none of chirality $-\chi$. Then,
the electromagnetic current $j_{\chi}^{\mu}$ of these electrons in
the presence of electromagnetic fields $\boldsymbol{E}$ and $\boldsymbol{B}$
would satisfy ($e>0$ is the unit electric charge and $\hbar$ is
the reduced Planck's constant)
\begin{equation}
\partial_{\mu}j_{\chi}^{\mu}=-\chi\frac{e^{3}}{4\pi^{2}\hbar^{2}}\boldsymbol{E}\cdot\boldsymbol{B}\label{eq:chiral anomaly}
\end{equation}
i.e., charge would not be conserved! (\ref{eq:chiral anomaly}) can
equivalently be written in terms of the electromagnetic fields strength
$F_{\mu\nu}=\partial_{\mu}A_{\nu}-\partial_{\nu}A_{\mu}$, where $A_{\mu}$
is the vector potential, as
\begin{equation}
\partial_{\mu}j_{\chi}^{\mu}=-\chi\frac{e^{3}}{32\pi^{2}\hbar^{2}}\epsilon^{\mu\nu\rho\lambda}F_{\mu\nu}F_{\rho\lambda}\label{eq:chiral anomaly-1}
\end{equation}
where $\epsilon^{\mu\nu\rho\lambda}$ is the antisymmetric tensor.
(\ref{eq:chiral anomaly}) and (\ref{eq:chiral anomaly-1}) seem absurd;
however, they makes sense instantly when one recalls that in reality,
Weyl nodes always come in pairs of opposite chiralities and the total
current $j_{+}^{\mu}+j_{-}^{\mu}$ is therefore conserved. In fact,
the requirement of current conservation is an equally good argument
for why the total chirality of the Weyl nodes must vanish. Classically,
currents are always conserved no matter what the dispersion. Thus,
(\ref{eq:chiral anomaly}) is a purely quantum phenomenon and is an
upshot of the path integral for Weyl fermions coupled to an electromagnetic
field not being invariant under separate gauge transformations on
left-handed and right-handed Weyl fermions, even though the action
is. This will be explained in more detail in Sec. \ref{sec:The-chiral-anomaly}.

The purpose of this brief review is to recap some of the strange transport
phenomena associated with the chiral anomaly in WSMs that have been
discussed in the literature so far. The field is mushrooming, so we
make no attempt to be exhaustive. Instead, we describe results that
are relatively simple, experiment-friendly and \emph{firsts}, to the
best of our knowledge. This is an introductory review targeted mainly
towards readers new to the subject. Thus, the results are sketched
rather than expounded, and readers interested in further details of
any result are encouraged to follow up by consulting the original
work.

Before embarking on the review, we skim over another striking feature
of WSMs -- surface states known as \emph{Fermi arcs}. Although this
review does not focus on the Fermi arcs, they are such a unique and
remarkable characteristic of WSMs that it would be grossly unfair
to review WSMs without mentioning Fermi arcs.

Topological band structures are invariably endowed with topologically
protected surface states, and WSMs are no exception. The Fermi surface
of a WSM on a slab consists of unusual states known as Fermi arcs.
These are essentially a 2D Fermi surface; however, part of this Fermi
surface is glued to the top surface and the other, to the bottom.
On each surface, Fermi arcs connect the projections of the bulk Weyl
nodes of opposite chiralities onto the surface, as shown in Figure
\ref{fig:Weyl-semimetal} for the case of two Weyl nodes. A simple
way to understand the presence of Fermi arcs is by recalling that
momentum space slices not containing Weyl nodes are Chern insulators
whose Chern numbers change by unity as one sweeps the slices past
a Weyl node. Thus, if the slices far away from the nodes have a Chern
number of $0$, i.e., the insulators are trivial, the slices between
the Weyl nodes are all Chern insulators with unit Chern number. The
Fermi arcs, then, are simply the edge states of these insulators.
Once WSMs with clean enough surfaces are found, the Fermi arcs should
be observable in routine photoemission experiments.

\begin{figure}
\begin{centering}
\includegraphics[width=0.3\columnwidth]{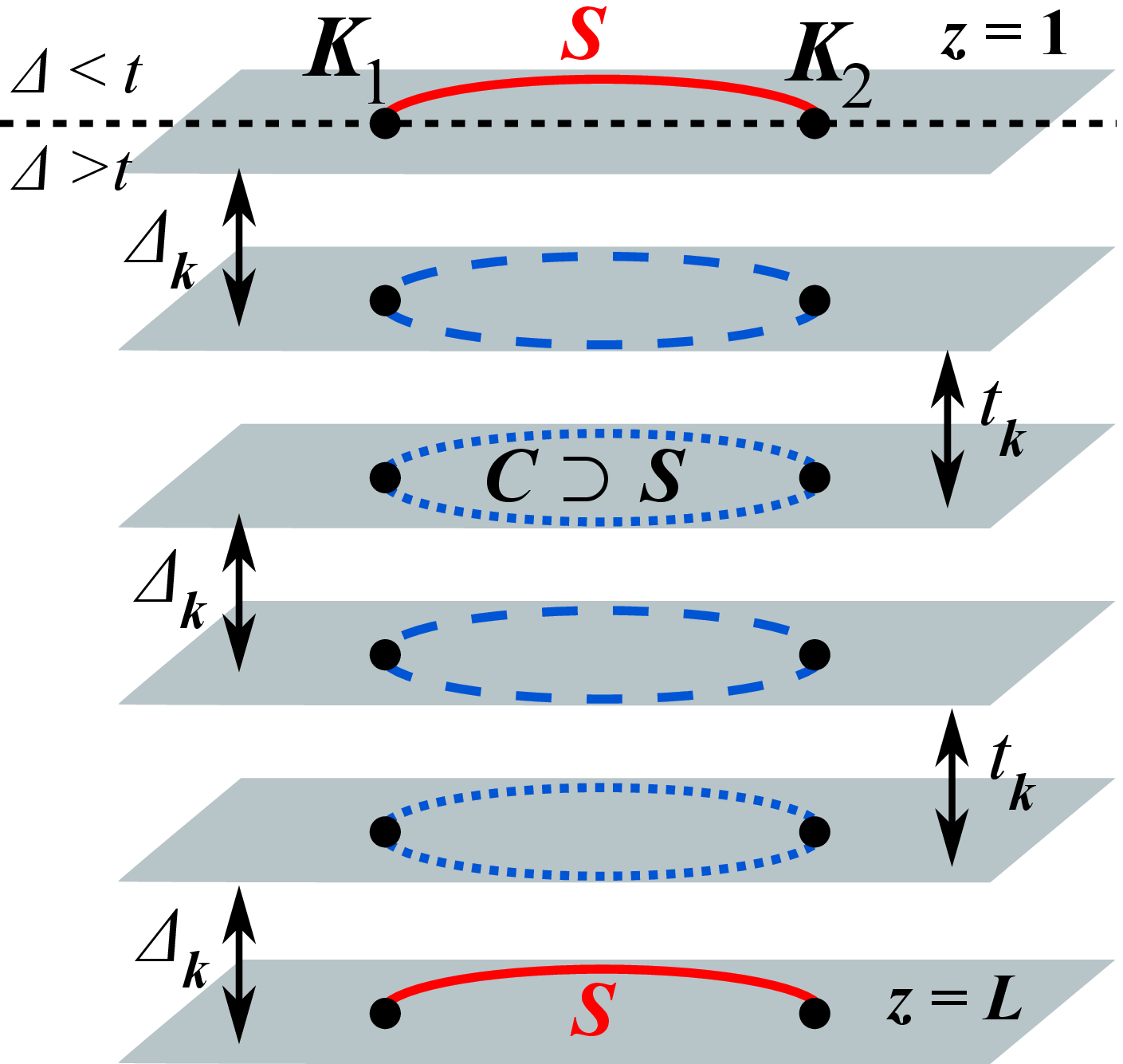}~~\includegraphics[width=0.3\columnwidth]{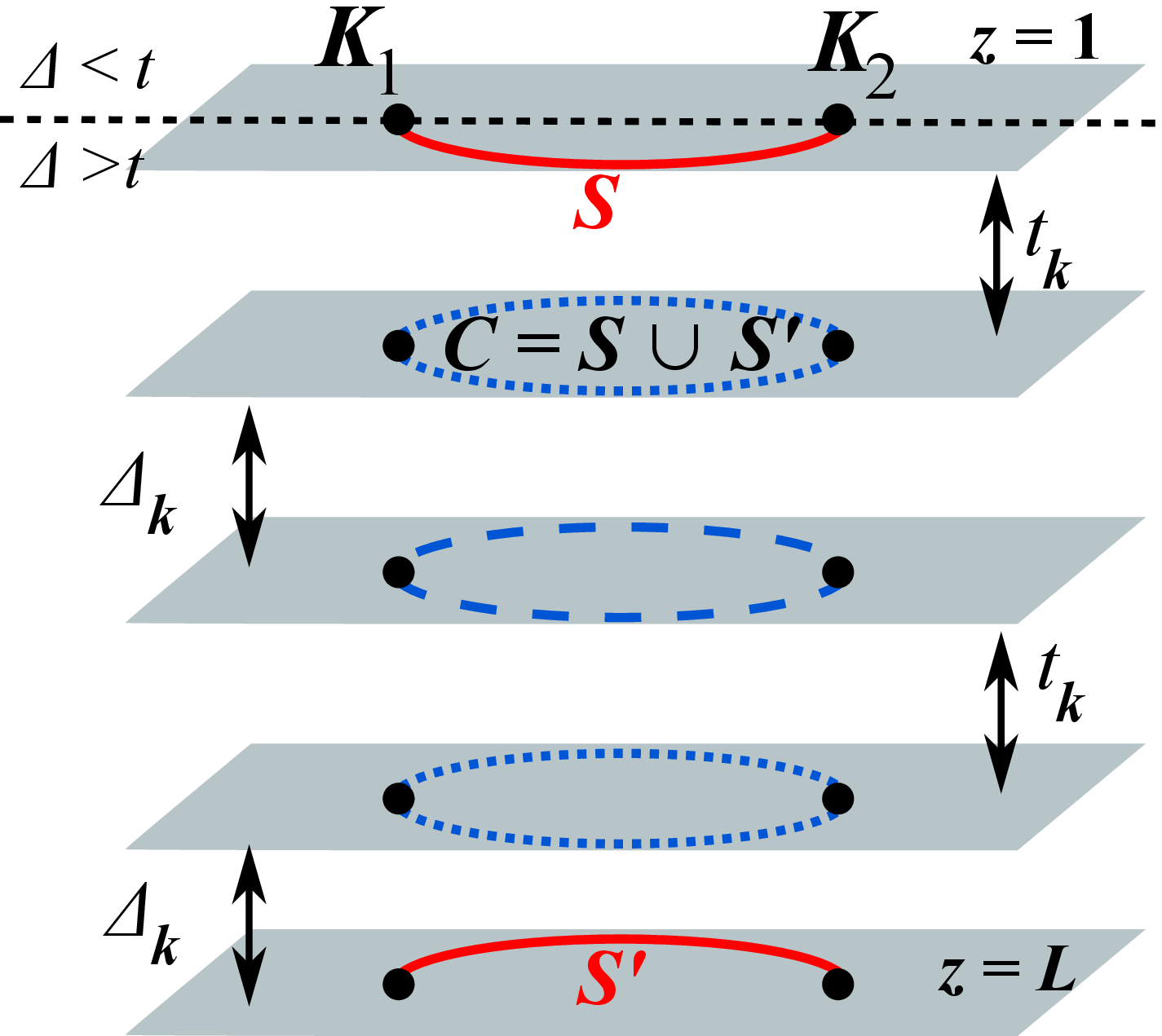}
\par\end{centering}

\caption{Fermi arcs as the residual states of the process of gapping out a
stack of 2D Fermi surfaces with momentum-dependent interlayer tunneling.
Dotted (dashed) blue curves $C$ represent 2D electron (hole) Fermi
surfaces, solid red lines on the end layers labeled $S$ or $S^{\prime}$
denote Fermi arcs, and $\Delta_{\boldsymbol{k}}$ and $t_{\boldsymbol{k}}$
are momentum-dependent interlayer hopping amplitudes whose relative
magnitudes change at the black dots $\boldsymbol{K}_{1}$ and $\boldsymbol{K}_{2}$
as one moves along the Fermi surface in any layer. Changing the boundary
conditions in the left figure by peeling off the topmost layer gives
the right figure, which has a different Fermi arc structure. (\citet{HosurFermiArcs})
\label{fig:Fermi-arcs-layering}}
\end{figure}

Alternately, Fermi arcs can also be understood as the states left
behind by gapping out a stack of 2D Fermi surfaces by interlayer tunneling
in a chiral fashion (\citet{HosurFermiArcs}). In particular, consider
a toy model consisting of a stack of alternating electron and hole
Fermi surfaces. For short-ranged interlayer tunneling, each point
on each Fermi surface can hybridize in two ways -- either with a state
in the layer above or with a state in the layer below. If the interlayer
tunneling is momentum dependent, such that the preferred hybridization
is different for different parts of the Fermi surface, then the points
at which the hybridization preference changes become Weyl nodes in
the bulk while the end layers have leftover segments that do not have
partners to hybridize with and survive as the Fermi arcs. This is
shown in Fig \ref{fig:Fermi-arcs-layering}. In this picture, the
topological nature of the Fermi arcs is not apparent; however, the
way they connect projections of Weyl nodes on the surface becomes
transparent. This is in contrast to the previous description of Fermi
arcs, in which figuring out what boundary conditions correspond to
what connectivity for the Fermi arcs is a highly non-trivial task.
The layering picture gives a systematic way to generate Fermi arcs
of the desired shape and connectivity, thus facilitating theoretical
studies of Fermi arcs significantly.

The rest of the review is organized as follows. We begin by recapping
electric transport in WSMs, which characterizes the linear dispersion
and not the chiral anomaly \emph{per se}, in Sec. \ref{sec:Electric-transport}.
This is followed by an intuitive explanation for the anomaly in Sec.
\ref{sec:Poor-man's-approach}. A more formal derivation of the anomaly
ensues in Sec. \ref{sec:The-chiral-anomaly}, and is succeeded by
a description of several simple but striking transport experiments
that can potentially serve as signatures of the anomaly in Sec. \ref{sec:Anomaly-induced-magnetotransport}.
We conclude with a discussion of the systems in which WSMs have been
predicted and the promise the field of WSMs and gapless topological
phases holds.

\section{Electric transport -- bad metal or bad insulator?\label{sec:Electric-transport}}

The optical conductivity of metals is characterized by a zero frequency
Drude peak, whose width is determined by the dominant current relaxation
process, in the real, longitudinal part. The Drude peak appears because
a metal has gapless excitations which carry current and generically,
have non-zero total momentum of the electrons. In the DC limit, relaxation
from a current-carrying state to the ground state typically involves
electrons scattering off of impurities or, at finite temperatures,
interactions with phonons. Each scattering processes produces its
own characteristic temperature ($T$) dependence -- the disorder dependent
DC conductivity is $T$ independent as long as the impurities are
dilute and static, while the rate of inelastic scattering off of phonons
grows rapidly with temperature, giving rise to a $T^{5}$ dependence
of the DC resistivity. Importantly, both these processes, besides
relaxing the current, relax the total electron momentum as well. On
the other hand, electron-electron interactions conserve momentum and
cannot contribute to the conductivity.

Band insulators, on the other hand, have a vanishing DC conductivity
simply because they have a band gap, but show a bump in the optical
conductivity when the frequency becomes large enough to excite electrons
across the gap. A similar bump occurs in the temperature dependence
as well when electrons can be excited across the gap thermally. Weak
disorder does not change either behavior significantly.

How do WSMs behave? They obviously must rank somewhere between metals
and insulators. But are they better thought of as conductors with
a vanishingly small density of states at the Fermi level, or as insulators
with a vanishing band gap? This question was addressed recently, in
the continuum limit (\citet{HosurWeylTransport,BurkovNodalSemimetal})
as well as using a lattice model with eight Weyl nodes (\citet{RosensteinTransport}).

\citet{HosurWeylTransport} showed that WSMs, like graphene (\citet{FritzGrapheneConductivity})
and 3D Dirac semimetals (\citet{GoswamiDiracTransport}), actually
exhibit a phenomenon that neither metals nor insulators do -- DC transport
driven by Coulomb interactions between electrons alone, even in a
clean system. This is because all these systems possess a particle-hole
symmetry about the charge neutrality points, at least to linear order
in deviations from these points in momentum space. As a result, there
exist current carrying states consisting of electrons and holes moving
in opposite directions with equal and opposite momenta. Since the
total momentum is zero, Coulomb interactions can indeed relax these
states. A quantum Boltzmann calculation gives the DC conductivity
at temperature $T$
\begin{equation}
\sigma_{dc}(T)=\frac{e^{2}}{h}\frac{k_{B}T}{\hbar v_{F}(T)}\frac{1.8}{\alpha_{T}^{2}\log\alpha_{T}^{-1}}\label{eq:DCcoulomb}
\end{equation}
where $v_{F}(T)$ and $\alpha_{T}$ are the Fermi velocity and fine
structure constant renormalized (logarithmically) to energy $k_{B}T$.
(\ref{eq:DCcoulomb}) can be understood within a picture of thermally
excited electrons diffusively, as follows. Einstein's relation $\sigma_{dc}(T)=e^{2}D(T)\frac{dn(T)}{d\mu}$
expresses $\sigma_{dc}$ in terms of the density of states at energy
$k_{B}T$, $\frac{dn(T)}{d\mu}\sim\frac{(k_{B}T)^{2}}{(\hbar v_{F})^{3}}$,
and the diffusion constant $D(T)=v_{F}^{2}\tau(T)$, where $\tau(T)$
is the temperature dependent transport lifetime. Now, the scattering
cross-section for Coulomb interactions must be proportional to $\alpha^{2}$
because scattering matrix elements are proportional to $\alpha$.
Since $T$ is the only energy scale in the problem, $\tau^{-1}(T)\sim\alpha^{2}T$
on dimensional grounds. This immediately gives (\ref{eq:DCcoulomb})
upto logarithmic factors. This behavior has already been seen approximately
in the pyrochlore iridates Y$_{2}$Ir$_{2}$O$_{7}$ (\citet{WeylResistivityMaeno}),
Eu$_{2}$Ir$_{2}$O$_{7}$ under pressure (\citet{EuIridateExperiments})
and Nd$_{2}$Ir$_{1-x}$Rh$_{x}$O$_{7}$ ($x\approx0.02\mbox{-}0.05$)
(\citet{UedaNd2Ir2O7}), all of which are candidate WSMs.

\begin{figure}
\begin{centering}
\includegraphics[width=0.6\columnwidth]{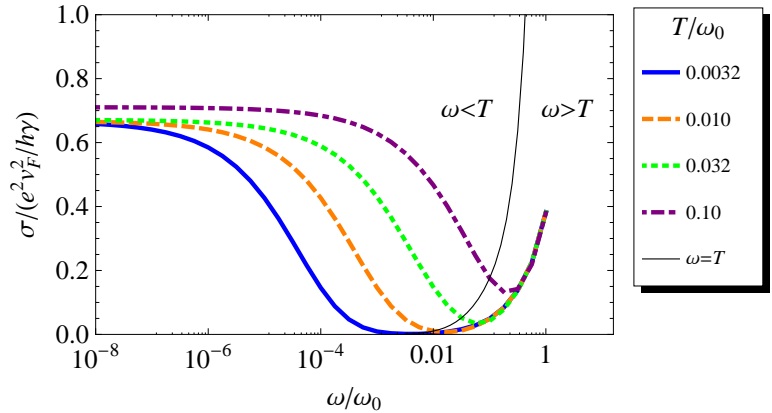}
\par\end{centering}

\caption{Linear-log plot of the optical conductivity of a WSM with Gaussian
disorder computed within a Born approximation. The disorder strength
is characterized by $\gamma$ (or $\omega_{0}=2\pi v_{F}^{3}/\gamma$).
(\citet{HosurWeylTransport})\label{fig:disorder conductivity}}
\end{figure}

On the other hand, non-interacting WSMs with chemical potential disorder
act like bad metals in some of their transport properties (\citet{HosurWeylTransport,BurkovNodalSemimetal,WeylMultiLayer}).
Fig \ref{fig:disorder conductivity} shows the temperature and frequency
dependence of the optical conductivity of disordered WSMs. Like metals,
it has a Drude peak whose height is set by the disorder strength.
However, the width of the peak goes as $T^{2}$, unlike metals where
the peak width has a weaker dependence -- $\sqrt{T}$ -- on temperature.
Thus, the conductivity at small non-zero frequency falls faster as
the temperature is lowered in WSMs as compared to ordinary metals.
At high frequencies, the behavior is entirely different. At $\hbar\omega\gg k_{B}T$,
the conductivity grows linearly with the frequency: $\sigma_{xx}(\omega)=\frac{e^{2}}{12h}\frac{\omega}{v_{F}}$
per Weyl node. This is expected from dimensional analysis since the
only physical energy scale under these circumstances is the frequency.
Importantly, such behavior is unparalleled in metals or insulators.

In summary, WSMs are neither metallic nor insulating in most of their
electric transport properties. However, if one is forced to put a
finger on which of the two more common phases they behave like, (bad)
`metals' is more accurate than (bad) `insulators'.

\section{The chiral anomaly -- poor man's approach\label{sec:Poor-man's-approach}}

We now turn to the main focus of this review -- the chiral anomaly
and related anomalous magnetotransport. This is where the story really
starts to get fascinating and WSMs start displaying a slew of exotic
properties unheard of in conventional electronic phases. To start
off, we present a quick caricaturistic derivation of the anomaly to
give the reader a feel for the microscopic physics that is at play
here.

\begin{figure}
\begin{centering}
\includegraphics[width=0.3\columnwidth]{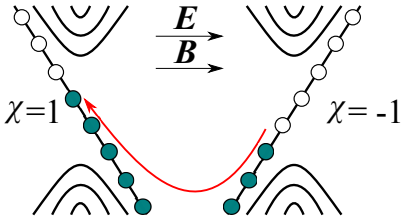}
\par\end{centering}

\caption{Charge pumping between Weyl nodes in parallel electric and magnetic
fields in the quantum limit. Each point in the dispersions is a Landau
level. Filled (empty) circles denote occupied (unoccupied) states.
Only occupation of the chiral zeroth Landau levels are shown because
they are the only ones that participate in the pumping.\label{fig:Charge-pumping}}
\end{figure}

In a magnetic field $\boldsymbol{B}$, the spectrum of the Hamiltonian
for a single Weyl node consists of Landau levels of degeneracy $g=\frac{BA_{\perp}}{h/e}$,
where $A_{\perp}$ is the cross-section transverse to $\boldsymbol{B}$,
dispersing along $\boldsymbol{B}$. Crucially, the zeroth Landau disperses
only one way, the direction of dispersion depending on the chirality
of the Weyl node.
\begin{eqnarray}
\epsilon_{n} & = & v_{F}\mbox{sign}(n)\sqrt{2\hbar|n|eB+(\hbar\boldsymbol{k}\cdot\hat{\boldsymbol{B}})^{2}},\, n=\pm1,\pm2,\dots\nonumber \\
\epsilon_{0} & = & -\chi\hbar v_{F}\boldsymbol{k}\cdot\hat{\boldsymbol{B}}
\end{eqnarray}
Suppose the temperature and the chemical potential are much smaller
than $v_{F}\sqrt{\hbar eB}$. Then, only the zeroth Landau level is
relevant for the low energy physics and we are in the so-called ``quantum
limit''. If an electric field $\boldsymbol{E}$ is now applied in
the same direction as $\boldsymbol{B}$, all the states move along
the field according to $\hbar\dot{\boldsymbol{k}}=-e\boldsymbol{E}$.
The key point is that the zeroth Landau level is chiral, i.e., it
disperses only one way for each Weyl node. Therefore, motion of the
states along $\boldsymbol{E}$ corresponds to electrons disappearing
from right-moving band and reappearing in the left-moving one, as
depicted in Fig \ref{fig:Charge-pumping}. In other words, the charge
in each of the $g$ chiral Landau bands is non-conserved, and each
of these bands exhibits a \emph{1D chiral anomaly}, given by $\partial Q_{\chi}^{1D}/\partial t=e\chi L_{B}|\dot{\boldsymbol{k}}|/2\pi=-e^{2}\chi L_{B}|\boldsymbol{E}|/h$,
where $L_{B}$ is the system size in the direction of $\boldsymbol{B}$.
Multiplying by $g$ gives the 3D result 
\begin{equation}
\frac{\partial Q_{\chi}^{3D}}{\partial t}=g\frac{\partial Q_{\chi}^{1D}}{\partial t}=-V\frac{e^{3}}{4\pi^{2}\hbar^{2}}\boldsymbol{E}\cdot\boldsymbol{B}
\end{equation}
where $V=A_{\perp}L_{B}$ is the system volume. This is the same as
(\ref{eq:chiral anomaly}) in the special case of a translationally
invariant system, in which the current due to a single Weyl node is
divergence free $\boldsymbol{\nabla}\cdot\boldsymbol{j}_{\chi}=0$.
The above ``quantum limit'' derivation is the simplest and most
intuitive way to understand (\ref{eq:chiral anomaly}). The result,
however, is not restricted to quantum limit. Indeed, charge pumping
between Weyl nodes of opposite chiralities was recently shown in a
purely semiclassical formalism as well, by \citet{SonSpivakWeylAnomaly}.

\section{The chiral anomaly -- general derivation\label{sec:The-chiral-anomaly}}

A more general field theoretic to understand the chiral anomaly is
by making the following observation. Just as there exists a quantum
Hall state in two dimensions which carries chiral 1D gapless edge
states, there is an analogous 4D state whose surface has chiral 3D
gapless edge states with opposite chiralities localized on opposite
surfaces (\citet{Zhang4DQH}). Now, in the absence of any internode
scattering, the two nodes of a WSM can be thought of as such surface
states, and the anomaly must emerge from the surface theory of the
hypothetical 4D quantum Hall state.

The bulk effective theory for electromagnetic fields for the 4D state
contains the topological third Chern-Simons term: 
\begin{equation}
S_{QH}^{4D}=\int\mathrm{d}^{5}x\left(\frac{e^{3}}{8\pi^{2}\hbar^{2}}\epsilon^{\mu\nu\rho\sigma\lambda}A_{\mu}\partial_{\nu}A_{\rho}\partial_{\sigma}A_{\lambda}+A_{\mu}j^{\mu}\right)
\end{equation}
just as that in a 2D quantum Hall state contains the second Chern
Simons term%
\footnote{It also contains the usual Maxwell term $S_{Max}\sim\int\mathrm{d}^{5}xF_{\mu\nu}F^{\mu\nu}$.
As we shall see in a moment, the anomaly stems from broken gauge invariance
of the Chern-Simons action at the boundary. $S_{Max}$ is gauge invariant
everywhere, so it does not contribute to the anomalous physics and
will thus be dropped. An important difference between 2+1D and 4+1D
Chern-Simons theories is that the former contains only one derivative
and hence, dominates the Maxwell term in the long-distance physics,
whereas the latter has two derivatives and competes with the Maxwell
term even at long distances.%
}. $S_{QH}^{4D}$ is well-defined in the bulk, but it violates gauge
invariance at the boundary. As a consequence, some of the gauge degrees
of freedom become physical and survive as gapless fermionic states
glued to the boundary, viz., the Weyl nodes. To see the anomaly in
this picture, let us generalize the procedure outlined by \citet{Maeda2DAnomaly}
in two dimensions and insert a step function $\Theta(x_{4})$ to simulate
a surface normal to $x_{4}$, the imaginary dimension, into the action:
\begin{equation}
\tilde{S}_{QH}^{4D}=\int\mathrm{d}^{5}x\left(\frac{e^{3}}{8\pi^{2}\hbar^{2}}\Theta(x_{4})\epsilon^{\mu\nu\rho\sigma\lambda}A_{\mu}\partial_{\nu}A_{\rho}\partial_{\sigma}A_{\lambda}+A_{\mu}j^{\mu}\right).
\end{equation}
The equation of motion $\frac{\partial\mathcal{L}}{\partial A_{\mu}}=\partial_{\nu}\left(\frac{\partial\mathcal{L}}{\partial_{\nu}A_{\mu}}\right)$
now contains an extra term on the right hand side proportional to
$\partial_{4}\Theta(x_{4})=\delta(x_{4})$, which is clearly localized
on the boundary. After an integration by parts, the boundary current
can be shown to satisfy (\ref{eq:chiral anomaly}). This is the essence
of the Callan-Harvey mechanism, which is well-known in the context
of the 2D integer quantum Hall effect but is equally well applicable
here. In this picture, the charge carried by the boundary states is
not conserved because it can always vanish into the bulk and reappear
on a different boundary.

A third approach to understanding the chiral anomaly entails encapsulating
the anomaly in the action itself rather than computing the chiral
current. The advantage of such an approach is that once the effective
action for the electromagnetic fields is known, physical transport
properties can be derived immediately. Below, we sketch one such technique
known as the Fujikawa rotation technique (see, for example, \citet{HosurRyuChiralTISC})
for deriving such an action. The technique transforms the action of
a WSM with two Weyl nodes into a massless Dirac action supplemented
by a topological $\theta$-term. The latter is a consequence of the
anomaly and is absent in an ordinary Dirac action. This was the approach
adopted in some recent works (\citet{ZyuninBurkovWeylTheta,GoswamiFieldTheory,ChenAxionResponse}).

Consider the continuum Euclidean action
\begin{equation}
S_{W}=\int\mathrm{d}^{4}x\bar{\psi}\gamma^{\mu}(\hbar i\partial_{\mu}-eA_{\mu}-b_{\mu}\gamma^{5})\psi
\end{equation}
where $\psi$ and $\bar{\psi}=\psi^{\dagger}\gamma^{0}$ are Grassman
spinor fields, $b_{\mu}$ is a constant 4-vector, $\gamma^{\mu},\,\mu=0\dots3$
are the standard $4\times4$ Dirac matrices and$\gamma^{5}=i\gamma^{0}\gamma^{1}\gamma^{2}\gamma^{3}$
is the chirality or ``handedness'' operator. In other words, right-handed
and left-handed Weyl nodes correspond to eigenstates of $\gamma^{5}$
with eigenvalues $+1$ and $-1$, respectively. All the five $\gamma$-matrices
anti-commute with one another. $S_{W}$ describes a Weyl metal (not
a WSM, unless $b_{0}=0$) with two Weyl nodes separated in momentum
space by $\boldsymbol{b}=(b_{x},b_{y},b_{z})$ and in energy by $b_{0}$
coupled to the electromagnetic field. $S_{W}$ is clearly invariant
under a chiral gauge transformation
\begin{equation}
\psi\to e^{-i\theta(x)\gamma^{5}/2}\psi\mbox{ or }\psi_{\pm}\to e^{\mp i\theta(x)/2}\psi_{\pm},\gamma^{5}\psi_{\pm}=\pm\psi_{\pm}\label{eq:chiralgt}
\end{equation}
which suggests that the chiral current $j_{ch}^{\mu}=e\bar{\psi}\gamma^{\mu}\gamma^{5}\psi$
is conserved. This is clearly wrong because we know that $j_{ch}^{\mu}$
conservation is violated according to (\ref{eq:chiral anomaly}).
What went wrong?

The flaw in the above argument becomes obvious when one realizes that
in a real condensed matter system, the right-handed and left-handed
Weyl nodes are connected at higher energies, so $\psi_{+}$ and $\psi_{-}$
cannot be gauge transformed separately as in (\ref{eq:chiralgt}).
In other words, the true action of the system changes under (\ref{eq:chiralgt}),
and the change comes from the regularization of the theory at high
energies.

To compute this change, let us perform such a transformation and see
what happens. If we choose $\theta(x)=(2\boldsymbol{b}\cdot\boldsymbol{r}-2b_{0}t)/\hbar$,
$b_{\mu}$ gets eliminated from $S_{W}$, leaving behind a massless
Dirac action $S_{D}=\int\mathrm{d}^{4}x\bar{\psi}i\gamma^{\mu}(\hbar\partial_{\mu}+ieA_{\mu})\psi$
in which both the chiral current $j_{ch}^{\mu}$ as well as the total
current $j^{\mu}=e\bar{\psi}\gamma^{\mu}\psi$ are truly conserved.
However, the measure of the path integral $\mathcal{Z}=\int\mathcal{D}\psi\mathcal{D}\bar{\psi}e^{-S_{W}[\psi,\bar{\psi}]}$
is not invariant under (\ref{eq:chiralgt}), which signals an anomaly.
More precisely, the Jacobian of the transformation is non-trivial
and can be interpreted as an additional term in the Dirac action:
\begin{equation}
\mathcal{D}\psi\mathcal{D}\bar{\psi}\to\mathcal{D}\psi\mathcal{D}\bar{\psi}\mathrm{det}\left[e^{i\theta(x)\gamma^{5}}\right]\equiv\mathcal{D}\psi\mathcal{D}\bar{\psi}e^{-S_{\theta}/\hbar}\implies S_{\theta}=-i\hbar\mathrm{Tr}\left[\theta(x)\gamma^{5}\right]\label{eq:Sthetadef}
\end{equation}
thus giving $S_{W}=S_{D}+S_{\theta}$.%
\footnote{Strictly speaking, the gauge transformation (\ref{eq:chiralgt}) to
remove $b_{\mu}$ from $S_{W}$ must be done in a series of infinitesimal
steps. However, the contribution to $S_{\theta}$ from each step happens
to be the same, so we are justified in doing the transformation at
once.%
} Information about the violation of chiral gauge invariance at high
energies is now expected to be contained in $S_{\theta}$.

The meaning of the trace in (\ref{eq:Sthetadef}) is highly subtle,
and it is not a simple trace of the matrix $\gamma^{5}$ multiplied
by a scalar function $\theta(x)$. If it were, $S_{\theta}$ would
vanish because $\gamma^{5}$ is traceless. Rather, it represents a
sum over a complete basis of fermionic states with suitable regularization.
A natural basis choice is the eigenstates of the Dirac operator $\slashed D=\gamma^{\mu}(\hbar\partial_{\mu}+ieA_{\mu})$;
a regularization method traditionally used in particle physics literature
is the heat kernel regularization, which exponentially suppresses
states at high energy. Thus,
\begin{eqnarray}
\mathrm{Tr}\left[\theta(x)\gamma^{5}\right] & = & \lim_{M\to\infty}\int\mathrm{d}^{4}x\sum_{n}\phi_{n}^{*}(x)e^{-\epsilon_{n}^{2}/M^{2}}\theta(x)\gamma^{5}\phi_{n}(x)\nonumber \\
 & = & \lim_{M\to\infty}\int\mathrm{d}^{4}x\sum_{n}\phi_{n}^{*}(x)e^{-\slashed D^{2}/M^{2}}\theta(x)\gamma^{5}\phi_{n}(x)\label{eq:trace for Stheta}
\end{eqnarray}
where
\begin{eqnarray}
\slashed D\phi_{n}(x) & = & \epsilon_{n}\phi_{n}(x)\nonumber \\
\int d^{4}x\phi_{n}^{*}(x)\phi_{m}(x)=\delta_{nm} & , & \sum_{n}\phi_{n}^{*}(x)\phi_{n}(y)=\delta(x-y)\label{eq:fermionbasis}
\end{eqnarray}
The right hand side in (\ref{eq:trace for Stheta}) can be evaluated
by Fourier transforming to momentum space and using the completeness
relations in (\ref{eq:fermionbasis}) (See \citet{ZyuninBurkovWeylTheta}
or \citet{GoswamiFieldTheory} for details). The result is $S_{\theta}$
in terms of the electromagnetic fields: 
\begin{equation}
S_{\theta}=\frac{ie^{2}}{32\pi^{2}\hbar^{2}}\int\mathrm{d}^{4}x\theta(x)\epsilon^{\mu\nu\rho\lambda}F_{\mu\nu}F_{\rho\lambda}=\frac{ie^{2}}{4\pi^{2}\hbar^{2}}\int\mathrm{d}^{4}x\theta(\boldsymbol{r},t)\boldsymbol{E}\cdot\boldsymbol{B}\label{eq:theta-term}
\end{equation}
Note that electromagnetic fields entered the derivation via the Dirac
operator $\slashed D$ in the regularization step in (\ref{eq:Sthetadef}).
One could choose a different regularization; a natural choice for
condensed matter systems would be to add non-linear terms to the dispersion.
However, electromagnetic fields will still enter the derivation via
minimal coupling and the final result for $S_{\theta}$ should be
the same. In fact, it turns out that some sort of regularization is
unavoidable, because the right hand side of (\ref{eq:Sthetadef})
without it is the difference of two divergent terms and is thus ill-defined.
We had earlier anticipated precisely this fact -- the chiral anomaly
is nothing but a shadow of the violation of chiral symmetry at high
energies in the low energy physics, and must originate in the high
energy regularization of the theory.

(\ref{eq:theta-term}) is eerily similar to the magnetoelectric term
that appears in the action of a 3D topological insulator. There is
an important difference, however. In topological insulators, $\theta(\boldsymbol{r},t)$
takes a constant value $\theta=\pi$ whereas here it is a spacetime
dependent scalar field. This subtle difference immediately leads to
novel topological properties in WSMs, some of which we discuss below.
Moreover, if translational symmetry is broken and the Weyl nodes are
gapped out by charge density wave order at the wavevector that connects
them, it can be shown that $\theta(\boldsymbol{r},t)$ survives as
phase degree freedom of the density wave, which is the Goldstone mode
of the translational symmetry breaking process (\citet{WeylCDW}).

\section{Anomaly induced magnetotransport\label{sec:Anomaly-induced-magnetotransport}}

Having understood the basic idea of the chiral anomaly, we now describe
several transport signatures of this effect.

\subsection{Negative magnetoresistance\label{sub:Negative-magnetoresistance}}

One of the first transport signatures of the chiral anomaly was pointed
out more than 30 years ago by Nielsen and Ninomiya (\citet{NielsenABJ}).
They noted that since Weyl nodes are separated in momentum space,
any charge imbalance created between them by an $\boldsymbol{E}\cdot\boldsymbol{B}$
field or otherwise requires large momentum scattering processes in
order to relax. In a sufficiently clean system, such processes are
relatively weak, resulting in a large relaxation time. An immediate
consequence of this is that the longitudinal conductivity along an
applied magnetic field, which is proportional to the relaxation time,
is extremely large. Moreover, a WSM in a magnetic field reduces to
a large number -- equal to the degeneracy of the Landau levels --
of decoupled 1D chains dispersing along the field as show in Fig \ref{fig:Charge-pumping}.
Therefore, the conductivity is proportional to the magnetic field
or, equivalently, the resistivity decreases with increasing magnetic
field. This phenomenon is termed as \emph{negative magnetoresistance}. 

\begin{figure}
\begin{centering}
\includegraphics[width=0.5\columnwidth]{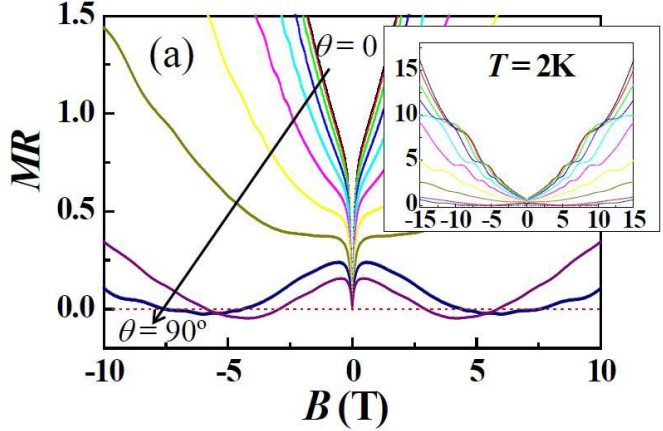}
\par\end{centering}

\caption{Magnetoresistance (MR) of Bi$_{1-x}$Sb$_{x}$ tuned to a quantum
critical point as a function of the magnetic field $B$. $\theta=90^{\circ}$
($\theta=0$) corresponds to longitudinal (transverse) magnetoresistance.
$B$ seemingly splits the Dirac node into two Weyl nodes in addition
to creating Landau levels. The initial rise in MR at very small fields
is attributed to weak antilocalization of the Dirac node, while the
chiral anomaly is responsible for the subsequent flattening or decrease
in it, as explained in Sec \ref{sub:Negative-magnetoresistance}.
The effect of the anomaly is clearly more pronounced for longitudinal
configurations. The upturn at very large fields is probably due to
localization of the 1D modes that constitute the quantum limit picture
described in the text. (Figure from \citet{BiSbKimMagnetoTransportExpt})\textbf{\label{fig:Magnetoresistance-BISb}}}
\end{figure}

Being one of the simplest signatures of the anomaly, negative magnetoresistance
has also been the first one to have been observed experimentally (\citet{BiSbKimMagnetoTransportExpt}).
The material on which the experiment was performed was Bi$_{0.97}$Sb$_{0.03}$.
Bi (Sb) is known to have topologically trivial (non-trivial) valence
bands in the sense of a 3D strong topological insulator. Therefore,
it is possible to fine tune Bi$_{1-x}$Sb$_{x}$ to the critical point
separating the two phases (\citet{FuKaneTIInversion,Murakami2007,HsiehTIDirac}).
The critical point has a single Dirac node in its bands structure.
\citet{BiSbKimMagnetoTransportExpt} applied a magnetic field at the
critical point, which not only created Landau levels but also split
the Dirac node into two Weyl nodes. The magnetoconductivity was subsequently
measured, the main result of which for our purposes is shown in Fig
\ref{fig:Magnetoresistance-BISb}. Beyond a very small field strength,
there is a clear negative contribution to the resistivity that is
enormous for longitudinal fields but very small for transverse ones.
The positive contributions to the resistivity at very small fields
are attributed to weak anti-localization of the Dirac node at the
critical point, while those at very large fields are probably because
the large number of 1D modes dispersing along the field become independent
1D systems and can be easily localized.

\subsection{Anomalous Hall and chiral magnetic effects \label{sub:AHECME}}

The next set of effects we shall discuss are the anomalous Hall effect
(AHE) and the chiral magnetic effect (CME). Although the effects appear
to be quite different, their derivation using the field theory outlined
in Sec \ref{sec:The-chiral-anomaly} shows that they are simply related
by Lorentz transformation and are conveniently discussed together.

By carrying out an integration by parts in (\ref{eq:theta-term}),
dropping boundary terms and Wick rotating to real time, $S_{\theta}$
can be written as
\begin{equation}
S_{\theta}=-\frac{e^{2}}{8\pi^{2}\hbar}\int\mathrm{d}t\mathrm{d}\boldsymbol{r}\partial_{\mu}\theta\epsilon^{\mu\nu\rho\lambda}A_{\nu}\partial_{\rho}A_{\lambda}
\end{equation}
Varying with respect to the vector potential gives the currents $j_{\nu}=\frac{e^{2}}{4\pi^{2}\hbar}\partial_{\mu}\theta\epsilon^{\mu\nu\rho\lambda}\partial_{\rho}A_{\lambda}$.
Recalling that $\theta(x)=(2\boldsymbol{b}\cdot\boldsymbol{r}-2b_{0}t)/\hbar$
gives 
\begin{equation}
\boldsymbol{j}=\frac{e^{2}}{2\pi^{2}\hbar^{2}}\boldsymbol{b}\times\boldsymbol{E}+\frac{e^{2}}{2\pi^{2}\hbar^{2}}b_{0}\boldsymbol{B}\label{eq:ahecme}
\end{equation}

The first term in (\ref{eq:ahecme}) is the AHE in the plane perpendicular
to $\boldsymbol{b}$. This can be understood based on the picture
for Fermi arcs presented in the introduction. We quickly repeat the
argument here: each 2D slice of momentum space perpendicular to $\boldsymbol{b}$
can be thought of as an insulator with a gap that depends on the momentum
parallel to $\boldsymbol{b}$. Since Weyl nodes are sources of unit
Chern flux with a sign proportional to their chirality, the slices
between $\boldsymbol{k}=-\boldsymbol{b}/2$ and $\boldsymbol{k}=\boldsymbol{b}/2$
are Chern insulators with unit Chern number, while those outside this
region are trivial insulators. Each of these Chern insulators has
unit Hall conductivity and as a result, the WSM has a Hall conductivity
proportional to $\boldsymbol{b}$. This result has been derived in
several recent works, both in lattice as well as in continuum models
(\citet{ChenAxionResponse,FangChernSemimetal,GoswamiFieldTheory,RanQHWeyl,VafekDiracReview,WeylMultiLayer,ZyuninBurkovWeylTheta}),
and has been accepted unanimously.

The second term in (\ref{eq:ahecme}), known as the chiral magnetic
effect (CME), is subtler, and has created some controversy. It predicts
an equilibrium dissipationless current parallel to the magnetic field
if the two Weyl nodes are at different energies. \citet{ZhouSemiclassicalTransport}
obtained the same result in a semiclassical limit. The CME is (deceptively,
as we will see) easy to understand in the DC continuum limit: if the
Weyl nodes are at different energies, the chiral zeroth Landau level
states from the two nodes will have different occupations and their
currents will not cancel each other. 

However, as pointed out by \citet{VazifehEMResponse}, this seems
to be at odds with some basic results of band theory. In particular,
a DC magnetic field reduces the 3D WSM to a highly degenerate 1D system
dispersing along the field. The total current DC along the field is
\begin{equation}
j_{B}=\int_{k_{L}}^{k_{R}}g\frac{\partial\epsilon}{\partial k}\mathrm{d}k=g(\epsilon_{R}-\epsilon_{L})\label{eq:jB}
\end{equation}
where $k_{L}$ and $k_{R}$ are the momenta of the Fermi points of
the 1D system, $g=\frac{BA_{\perp}}{h/e}$ is the Landau level degeneracy
and $\epsilon(k)$ is the 1D dispersion. At equilibrium, $\epsilon_{R}=\epsilon_{L}$,
which implies $j_{B}$ vanishes. \citet{VazifehEMResponse} argued
that the CME was an artifact of linearizing the dispersion near the
Weyl nodes, and is in fact absent in a full lattice model. The CME
seems wrong for another reason too. If a state carries a net DC current
$\boldsymbol{J}$, then Ohmic power $\sim\boldsymbol{J}\cdot\boldsymbol{E}$
can be supplied to or extracted from it by applying an appropriate
electric field $\boldsymbol{E}$. But a ground state is already the
lowest energy state, so it is not possible to extract energy from
it. Therefore it cannot carry a DC current (\citet{BasarTriangleAnomaly}).

Soon after, though, this claim was countered by \citet{ChenAxionResponse}
who showed, by rederiving the CME in a lattice model, that while the
CME is unambiguously non-zero at finite momentum $\boldsymbol{q}$
and frequency $\omega$, its DC limit depends on the order of the
limits $\omega\to0$ and $\boldsymbol{q}\to0$. If $\omega\to0$ is
taken first, one obtains a static system in which the electrons are
always in equilibrium and the CME vanishes as predicted by Ref. \citet{VazifehEMResponse}.
However, the effect survives if the $\boldsymbol{q}\to0$ limit is
evaluated first and is precisely that predicted by \ref{eq:ahecme}.
In this case, the electrons are not in equilibrium except \emph{at
the limit}, and neither the band theory argument nor the Ohmic dissipation
argument presented above applies.

\subsection{Non-local transport\label{sub:Non-local-transport}}

\begin{figure}
\begin{centering}
\includegraphics[width=0.3\columnwidth]{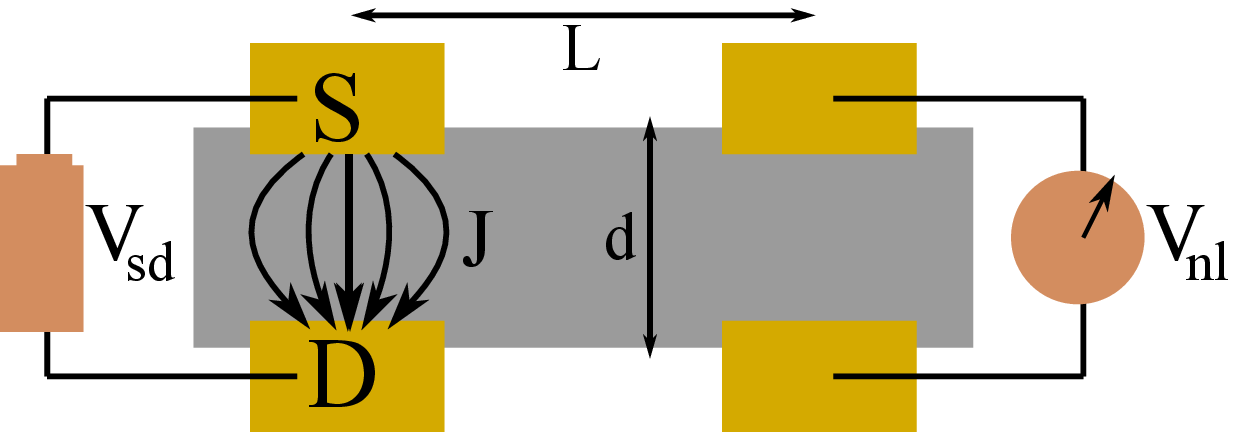}~~\includegraphics[width=0.5\columnwidth]{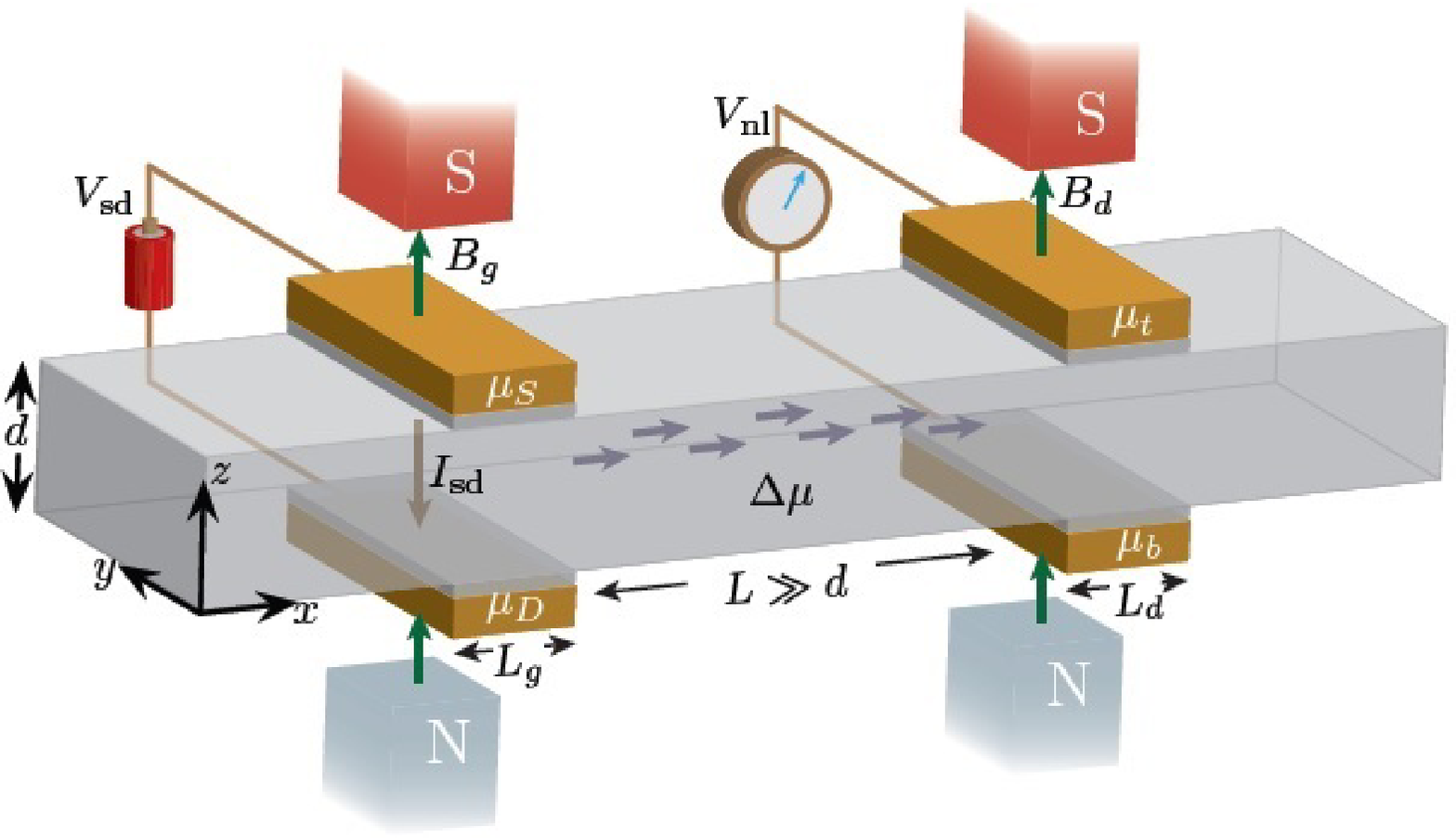}
\par\end{centering}

\caption{(Left) A current $J$ driven between the source ($S$) and drain ($D$)
leads in an ordinary piece of metal takes the path indicated by the
arrows inside the sample. The Ohmic voltage between the top and the
bottom surfaces decays over the length scale of the sample thickness
$d$ as one moves away from the injection region, so that $V_{nl}\sim V_{sd}e^{-L/d}$.
(Right) In a WSM, a local magnetic field $B_{g}$ along the injected
current generates a valley imbalance as a result of the anomaly. This
imbalance diffuses slowly if the intervalley scattering processes
are weak, and can be detected far away from the injection region by
a probe magnetic field $B_{p}$ which converts it into an electric
field. (Right figure from \citet{ParameswaranNonLocalTransport})\label{fig:nonlocaltransport}}
\end{figure}

Consider a 3D piece of ordinary metal with four contacts attached
to it, as shown in Fig \ref{fig:nonlocaltransport} (left). The two
contacts on the left, labeled source ($S$) and drain ($D$) inject
a current $J$ through the sample, whose typical path is indicated
by the arrows connecting the $S$ and the $D$ leads. As one moves
away from $S$ and $D$, the voltage drop between the upper and lower
surfaces falls because smaller and smaller segments of the current
lines are encountered by a path that goes vertically from the top
to the bottom surface. In particular, it can be shown that in the
``quantum limit'' where the mean free path is limited only by the
contacts, the nonlocal voltage drop $V_{nl}$ decays on the scale
of the sample thickness $d$. 

\citet{ParameswaranNonLocalTransport} showed that the situation in
WSMs in the presence of local magnetic fields is strikingly different.
As a consequence of the anomaly, a local magnetic field applied parallel
to the injected current generates an imbalance in the occupation numbers
of the two Weyl nodes locally. This imbalance diffuses over a length
scale $l$ determined by the rate of internode scattering processes
and hence, can be quite large -- even larger than the sample thickness
$d$. Thus, at distances $L\sim l\gg d$ there are no Ohmic voltages
but there exists a \emph{valley} imbalance, borrowing nomenclature
from semiconductor physics. Detecting the valley imbalance is challenging,
however, because it does not couple to electric fields. The anomaly
comes to the rescue, and the last piece of the puzzle entails applying
a local probe magnetic field. This couples to the valley imbalance
and produces a local electric field which can then be measured by
conventional methods.

\subsection{Chiral gauge anomaly\label{sub:Chiral-gauge-anomaly}}

The final effect we will describe is the response to a chiral gauge
field, i.e., a gauge field that has opposite signs for Weyl nodes
of opposite chirality. One way to create such a field is by applying
appropriate strains, as has been done in graphene (\citet{PeresGrapheneElectronicRev})
to apply opposite gauge fields to the two Dirac nodes there. More
generally, chiral gauge fields can be created in WSMs by exploiting
the fact that a general Weyl metal can be created from a Dirac semimetal
by breaking time-reversal and inversion symmetries. Then, fluctuations
in the perturbations that break these symmetries naturally act like
chiral gauge fields.

\citet{QiWeylAnomaly} pointed out that chiral gauge fields in a WSM
induce an anomaly not only in the chiral current but also in the total
electric current. The argument is simple. Suppose $A_{\mu}$ and $a_{\mu}$
are the electromagnetic and the chiral gauge fields, respectively,
and $F_{\mu\nu}$ and $f_{\mu\nu}$ the corresponding field strengths.
Then, a Weyl node of chirality $\chi$ feels a total gauge field $\mathcal{A}_{\mu}^{\chi}=A_{\mu}+\chi a_{\mu}$
and hence, suffers from an anomaly $\partial_{\mu}j_{\chi}^{\mu}=-\chi\frac{e^{3}}{4\pi^{2}\hbar^{2}}\epsilon^{\mu\nu\rho\lambda}\partial_{\mu}\mathcal{A}_{\nu}^{\chi}\partial_{\rho}\mathcal{A}_{\lambda}^{\chi}$
according to (\ref{eq:chiral anomaly-1}). Consequently, the conservation
laws for the chiral current $j_{ch}^{\mu}=j_{+}^{\mu}-j_{-}^{\mu}$
and the total current $j^{\mu}=j_{+}^{\mu}+j_{-}^{\mu}$ read
\begin{eqnarray}
\partial_{\mu}j_{ch}^{\mu} & = & -\frac{e^{3}}{16\pi^{2}\hbar^{2}}\epsilon^{\mu\nu\rho\lambda}\left(F_{\mu\nu}F_{\rho\lambda}+f_{\mu\nu}f_{\rho\lambda}\right)\label{eq:chiral chiral anomaly}\\
\partial_{\mu}j^{\mu} & = & -\frac{e^{3}}{8\pi^{2}\hbar^{2}}\epsilon^{\mu\nu\rho\lambda}F_{\mu\nu}f_{\rho\lambda}\label{eq:chiral gauge anomaly}
\end{eqnarray}
(\ref{eq:chiral chiral anomaly}) is the usual chiral anomaly which
receives additional contribution from the chiral gauge fields. (\ref{eq:chiral gauge anomaly}),
however, states that even the total current is not conserved if both
electromagnetic and chiral gauge fields are present. This can be rewritten
in terms of the chiral ``magnetic field'' $\boldsymbol{\beta}=\boldsymbol{\nabla}\times\boldsymbol{a}$
and chiral ``electric field'' $\boldsymbol{\varepsilon}=-\boldsymbol{\nabla}a_{0}-\frac{\partial\boldsymbol{a}}{\partial t}$
as
\begin{equation}
\partial_{\mu}j^{\mu}=-\frac{e^{3}}{2\pi^{2}\hbar^{2}}\left(\boldsymbol{\beta}\cdot\boldsymbol{E}+\boldsymbol{\varepsilon}\cdot\boldsymbol{B}\right)\label{eq:chiral gauge anomaly-1}
\end{equation}

\begin{figure}
\begin{centering}
\includegraphics[width=0.3\columnwidth]{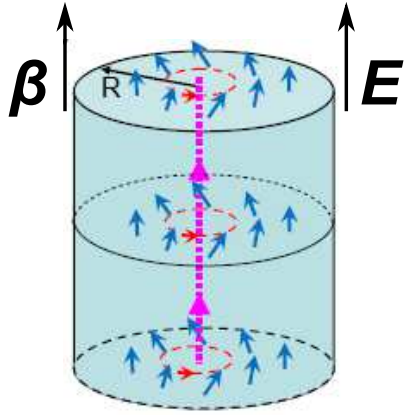}
\par\end{centering}

\caption{A vortex in the magnetization (small blue arrows) results in a chiral
magnetic field $\boldsymbol{\beta}$ along the vortex axis. An electric
field in the same direction then generates a one way current moving
along the vortex axis. (\citet{QiWeylAnomaly})}
\end{figure}

If time-reversal symmetry in the system is broken by ferromagnetism,
then $\boldsymbol{a}$ is proportional to the magnetic moment and
the chiral magnetic field corresponds to a ferromagnetic vortex. The
first term above, then, predicts a chiral current propagating along
the vortex axis. In the quantum limit picture of Sec \ref{sec:Poor-man's-approach},
this current is carried by zeroth Landau levels states, which now
disperse in the same direction for both Weyl nodes. An immediate question
is, how can charge not be conserved in a real system? The answer comes
from the realization that unlike $A_{\mu}$, $a_{\mu}$ is a physical
field and must be single valued. Since $a_{\mu}=0$ in vacuum, the
total flux $\boldsymbol{\beta}=\boldsymbol{\nabla}\times\boldsymbol{a}$
must vanish in a finite system. Equivalently, the total winding number
of all the vortices must be zero. Thus, there are equal numbers of
chiral and antichiral modes so that the total current is, in fact,
conserved. These modes are in different spatial regions, so the gauge
anomaly (\ref{eq:chiral gauge anomaly}) is still meaningful locally.

The second term is generated by a time-dependent magnetization, and
describes charge and current modulations in response to magnetization
fluctuations in the presence of $\boldsymbol{B}$. In other words,
it describes a coupling between plasmons (charge fluctuations) and
magnons (magnetic fluctuations), which is not present in ordinary
metals or semimetals.

\section{Materials realizations}

Theorists have occasionally enthused over 3D Dirac (\citet{Abrikosov})
and Weyl (\citet{NielsenFermionDoubling1,NielsenFermionDoubling2,NielsenABJ})
band structures for decades and it has long been known that the A-phase
of superfluid Helium-3 has Weyl fermions (\citet{VolovikBook,MengWeylSC}).
However, interest in them has recently been rekindled by the prediction
that some simple electronic systems realize them.

The first materials prediction was in the pyrochlore iridates family
-- R$_{2}$Ir$_{2}$O$_{7}$ -- where `R' is a rare earth element
(\citet{PyrochloreWeyl,KrempaWeyl,ChenIridate}). These candidate
WSMs are inversion symmetric but break time-reversal symmetry via
a special kind of anti-ferromagnetic ordering -- the `all-in/all-out'
ordering -- in which all the spins on a given tetrahedron point either
towards the center or away from it, and the ordering alternates on
adjacent tetrahedra. Available transport data on these materials are
roughly consistent with linearly dispersing bands (\citet{WeylResistivityMaeno,EuIridateExperiments,UedaNd2Ir2O7});
however, the evidence is far from conclusive as yet. In its footsteps
followed several proposals to engineer WSMs in topological insulators
heterostructures. Alternately stacking topological and ferromagnetic
insulators was shown to produce inversion symmetric WSMs in a certain
parameter regime with Weyl nodes separated along the stacking direction
(\citet{WeylMultiLayer}), while replacing the ferromagnetic insulators
with trivial, time-reversal symmetric ones and applying an electric
field perpendicular to the layers could give time-reversal symmetric
WSMs (\citet{HalaszWeyl}). A third option is to magnetically dope
a quantum critical point separating a topological and trivial insulator
(\citet{ChoTItoWeyl}). An advantage of this realization, as we saw
in Sec \ref{sub:Chiral-gauge-anomaly}, is that magnetization can
be used as a handle to dynamically modify the band structure; in return,
magnetic textures and fluctuations are endowed with physical properties
that uniquely characterize the topological nature of the underlying
bands (\citet{QiWeylAnomaly}). 

Some closely related phases have been predicted in real materials
as well. A ferromagnetic spinel HgCr$_{2}$Se$_{4}$ has been predicted
to form a double WSM, i.e., a WSM in which the Weyl nodes have magnetic
charge of $\pm2$ (\citet{FangChernSemimetal}). Passing light through
a cleverly designed photonic crystal gives it a dispersion that can
be fine-tuned to have with ``Weyl'' line nodes, i.e., a pair of
non-degenerate bands intersecting along a line (\citet{PhotonicCrystalWeyl}).
The surface states of such a crystal has flat bands, implying photon
states with zero velocity. Unlike electronic systems, such a band
structure can be obtained while preserving time-reversal and inversion
symmetries because there is no Kramers degeneracy for photons. Breaking
these symmetries reduces the line node to Weyl points. Finally, ab
initio calculations have predicted dispersions with degenerate Weyl
nodes in $\beta$-cristobalite BiO$_{2}$ (\citet{YoungDiracSemimetal}),
A$_{3}$Bi where A=Na or K (\citet{WangA3Bi}) and Cd$_{3}$As$_{2}$
(\citet{WangCd3As2}). Whereas the first two are inversion symmetric
Dirac semimetals, a combination of broken inversion symmetry and unbroken
crystal symmetries in Cd$_{3}$As$_{2}$ holds Weyl nodes of opposite
chiralities degenerate but gives them distinct Fermi velocities.

\section{Summary}

We have made a humble attempt at introducing Weyl semimetals and recapping
the recent theoretical and experimental developments in the transport
studies of this phase. The basics of WSMs can be summarized in three
main points. These points are interrelated and any one can be deduced
from any other.
\begin{itemize}
\item The first is the definition itself, that it has a band structure with
non-degenerate bands intersecting at arbitrary points in momentum
space. An immediate consequence of such band intersections is that
Weyl points are topologically robust as long as translational symmetry
is present. Moreover, the dispersion in the neighborhood of these
points is linear. We reviewed the transport properties contingent
on the linear dispersion in Sec \ref{sec:Electric-transport}. There,
we stated that unlike metals, WSMs can have a finite DC conductivity
driven purely by electron-electron interactions. We concluded the
section by stating that WSMs resemble metals more than insulators
based on disorder dependent transport.
\item The second key feature of WSMs is the chiral anomaly. Weyl nodes can
be characterized by a chirality quantum number of $\pm1$; the chiral
anomaly says that chiral charge, i.e., the number of quasiparticles
around Weyl nodes of fixed chirality, is not conserved in the presence
of parallel electric and magnetic fields. In other words, an $\boldsymbol{E}\cdot\boldsymbol{B}$
field pumps charge between Weyl nodes of opposite chiralities. We
presented two detailed derivations of the anomaly in Sec \ref{sec:The-chiral-anomaly}.
The first treats Weyl nodes as the surface states of a 4D quantum
Hall system. In this picture, the anomaly is understood as charge
pumping between opposite surfaces of a topological phase. The second
works entirely in three dimensions, and captures the anomaly in a
term in the action that supplements the one that gives the Weyl nodes
a linear dispersion. In this derivation, it becomes apparent that
the anomaly exists because separate gauge transformations on Weyl
fermions of opposite chiralities is forbidden by states at higher
energies.\\
Once the anomaly was introduced in some detail, we presented several
recent theoretical predictions for anomalous transport. These included
a negative contribution to magnetoresistance that has nothing to do
with weak localization (Sec \ref{sub:Negative-magnetoresistance}),
anomalous Hall effect as well as a current along an applied magnetic
field (Sec \ref{sub:AHECME}), extremely slow decay of a voltage in
the presence of a magnetic field in the same direction as the voltage
(Sec \ref{sub:Non-local-transport}), and non-conservation of ordinary
current due to fluctuations in the background fields that split a
Dirac node into Weyl nodes thus giving a WSM (Sec \ref{sub:Chiral-gauge-anomaly}).
We also flashed experimental results on Bi$_{1-x}$Sb$_{x}$ which
shows striking negative magnetoresistance consistent with the chiral
anomaly.
\item The third key characteristic of WSMs is peculiar surface states known
as Fermi arcs. These can either be thought of as the edge states of
Chern insulators layered in momentum space, or as the remnants of
the process of destroying a stack of electron and hole Fermi surfaces
in a chiral fashion. Once suitable materials are found, these states
should be easily observable in photoemission experiments.
\end{itemize}
We wrapped up the review by touching upon several systems in which
WSMs have been predicted to occur. So far, these include certain pyrochlore
iridates, heterostructures based on topological insulators and systems
with Dirac nodes perturbed by suitable symmetries. Bi$_{0.97}$Sb$_{0.03}$
placed in a magnetic field belongs to the last category. Magnetotransport
data on it is the best evidence thus far of the WSM phase being realized
in a real system. Nonetheless, the naturalness of 3D band crossings
and the number of materials predictions that have been made in a short
time are compelling indications that the field of WSMs and gapless
topological phases is well and truly blossoming.

\section{Acknowledgments}

We are indebted to Jerome Cayssol for inviting us to write the review.
We would like to thank Daniel Bulmash and Ashvin Vishwanath for valuable
feedback on the manuscript and the David and Lucile Packard Foundation
for financial support.

\bibliographystyle{plainnat}
\bibliography{C:/Users/Pavan/Dropbox/CurrentProjects/references}

\end{document}